# Extraction of the short-range defect potential parameters from available experimental data on the graphene resistance

*N.E. Firsova. S.A Ktitorov*

*Ioffe Institute, St. Petersburg, Russia*

We consider a problem of obtaining information about the scattering potentials of the monolayer graphene sample using available experimental data on its resistance. We have in mind a development of the study describing super-high mobility electrons in suspended samples without chemical doping. As far as practical absence of the doping impurities in this case makes the Coulomb scattering negligible, we consider models of the short-range scattering potentials. The model of short-range potential is assumed to be supported by the close vicinity of the ring or the circumference of a circle. The diameter of circles is supposed to be of the order of the crystal lattice spacing. The empty core of the model potential guarantees the suppression of nonphysical shortwave modes.

Two models are investigated: the delta function on the circumference of a circle (delta shell) and the annual well. An advantage of the former is simplicity, while a virtue of the latter is regularity. We consider scattering of electrons by these potentials and obtain exact explicit formulae for the scattering data. We here discuss application of these formulae for calculation of observables. Namely, we analyze the contribution of this scattering into the graphene resistance and plot the resistivity as a function of the Fermi energy according to our theoretical formulae. The obtained results are consistent with experiment, where the resistance was measured as a function of the Fermi momentum on the suspended annealed graphene. This fact gives a possibility to find parameters of the modeled potential on the base of the available experimental data on resistance of the suspended graphene sample with the gate voltage controlled Fermi level position. It is clear to be very important for applications.

## 1. Introduction

In [1-3] we investigated the model of the short-range scattering potentials assumed to be supported by the circumference of a circle. The diameter of circles is supposed to be of the order of the crystal lattice spacing. Such delta function form of the model potentials with the empty core guarantees the suppression of nonphysical shortwave modes. To analyze this problem the two-dimensional Dirac equation was considered

$$\hat{H}\psi(x,y) \equiv \left[-i\hbar v_F \sum_{\mu=1}^{2}\hat{\sigma}_\mu \partial_{x_\mu} + \hat{V}\right]\psi(x,y) = E\psi(x,y) \qquad (1)$$

where $v_F$ is the Fermi velocity, $\hat{\sigma}_\mu$ are the Pauli matrices, $\psi(r)$ is the two-component spinor. In [1-3] the following potential of delta function type supported by the circumference of a circle was considered

$$V(r) = -V_0 \delta(r - r_0). \qquad (2)$$

In [2] the formula for resistance $R$ is exact in the limit of long waves (i.e. in vicinity of the Dirac point) was derived for an arbitrary magnitude of perturbation within the range of the Boltzmann equation validity.

$$\frac{R}{R_0} = \pi^2 \tilde{N}_i \left[\tan 2\tilde{V}_0\right]^2, \qquad (3)$$

Here $R$ is the resistance,

$$R_0 = \sigma_0^{-1}, \quad \sigma_0 = 4\frac{e^2}{h}, \quad \tilde{N}_i = r_0^2 N_i, \quad \tilde{V}_0 = \frac{V_0}{2\hbar v_F}, \qquad (4)$$

and $N_i$ is the defects concentration. In [1] the resistance was calculated using the Born approximation on the basis of results from [4] in the wide region of de Broglie wave lengths, but in the limit of the small magnitude of the defect potential:

$$\frac{R}{R_0} = 4\pi \tilde{N}_i \tilde{V}_0^2 I(k), \qquad (5)$$

where

$$I(k) = \int_0^\pi d\theta [\sin\theta]^2 J_0^2\left(2k\sin\left(\frac{\theta}{2}\right)\right). \qquad (6)$$

In [1] the plots of the theoretical formulae (3)-(6) for the resistance as well as experimental ones obtained in [5] are presented. Theoretical formulae proved to be rather close to the experimental curves. This consistency allowed us to extract from the experimental results some information about inner parameters. In particular in [1] from the derived theoretical formulae for the case of delta potential model in the vicinity of the Dirac point it was obtained

$$\frac{R_{max}}{R_0} = N_i \left(2\pi r_0 \tilde{V}_0\right)^2 = N_i \left(L_{circ} \tilde{V}_0\right)^2, \qquad (7)$$

where $L_{circ}$ is the circumference length, $R_{max}$ is a value of the resistance in the vicinity of the Dirac point, where the resistance has its maximum.

## 2. Calculation of resistance in case of the model supported by the annual well.

Let us consider now the case of the potential model supported by the annual well continuing the investigation began in [6]. This model is more realistic, than the delta shell. We shall calculate the resistance for this case in the Born approximation and describe, what information on the characteristic parameters can be extracted from the obtained results and experimental data [5] for the investigated graphene sample.

We analyze the equation (1) with the potential

$$V(r) = \begin{cases} -U_0 & \text{if } r_1 < r < r_2 \\ 0 & \text{otherwise} \end{cases} \qquad (8)$$

We will calculate the resistance in the Born approximation. For conductivity we have the well known formula

$$\sigma(E_F) = \sigma_0 \frac{E_F}{\hbar} \tau, \quad \tau = \frac{1}{\Lambda_{tr} v_F N_i}, \quad \sigma_0 = 4\frac{e^2}{h}$$

where $\tau$ is the relaxation time, $\Lambda_{tr}$ is the transport cross section. We rewrite this formula as follows

$$\sigma(E_F) = \sigma_0 \frac{E_F}{\hbar v_F N_i} \frac{1}{\Lambda_{tr}} \qquad (9)$$

So we have a formula for resistance

$$R = R_0 \left(\frac{\hbar v_F N_i}{E_F}\right) \Lambda_{tr}, \qquad (10)$$

where the cross section reads [4]:

$$\Lambda_{tr} = \int_0^\pi d\theta(1-\cos\theta)|f(\theta)|^2,$$  (11)

and the Born amplitude

$$f(\theta) = -\frac{1}{\hbar v_F}\sqrt{\frac{k_F}{8\pi}}U_q\left(1+e^{-i\theta}\right),$$  (12)

$$U_q = \int d^2 r e^{-i\mathbf{qr}}V(r).$$  (13)

$V(r)$ is defined in (8) and

$$q = 2p\sin(\theta/2), \quad \theta = \text{angle}\,(\mathbf{k}_F,\mathbf{k}_F'),$$  (14)

Consequently

$$U_q = U_0\int_{r_1}^{r_2}dr\,r\int_0^{2\pi}d\varphi\,e^{-iqr\cos\varphi} = 2\pi U_0\int_{r_1}^{r_2}dr\,rJ_0(qr) = 2\pi U_0\int_{r_1}^{r_2}dr\,rJ_0\left(2k_F r\sin\frac{\theta}{2}\right)$$  (15)

Substituting this expression into (12) and then into (11) we obtain

$$\Lambda_{tr} = \frac{1}{(\hbar v_F)^2}\frac{k_F}{8\pi}(2\pi U_0)^2\int_0^\pi d\theta(1-\cos\theta)\left(1+e^{-i\theta}\right)\left(1+e^{i\theta}\right)\times$$

$$\times\left[\int_{r_1}^{r_2}dr\,rJ_0\left(2k_F r\sin\frac{\theta}{2}\right)\right]^2$$  (16)

As soon as

$$(1-\cos\theta)\left(1+e^{-i\theta}\right)\left(1+e^{i\theta}\right) = 2\sin^2\theta,$$

we have from (16)

$$\Lambda_{tr} = \frac{\pi k_F}{(\hbar v_F)^2}(U_0)^2 \int_0^\pi d\theta \sin^2\theta \left[\int_{r_1}^{r_2} dr\, rJ_0\left(2k_F r \sin\frac{\theta}{2}\right)\right]^2$$

Let us make a substitution: $\tilde{r} = k_F r$, $\tilde{r}_{1,2} = k_F r_{1,2}$ Then we obtain:

$$\Lambda_{tr} = \frac{\pi}{(\hbar v_F)^2 k_F^3}(U_0)^2 \int_0^\pi d\theta \sin^2\theta \left[\int_{\tilde{r}_1}^{\tilde{r}_2} d\tilde{r}\, \tilde{r}J_0\left(2\tilde{r} \sin\frac{\theta}{2}\right)\right]^2 \qquad (17)$$

Since $k_F = E_F/(\hbar v_F)$, we have:

$$\Lambda_{tr} = \frac{\pi(\hbar v_F)}{E_F^3}(U_0)^2 \int_0^\pi d\theta \sin^2\theta \left[\int_{\tilde{r}_1}^{\tilde{r}_2} d\tilde{r}\, \tilde{r}J_0\left(2\tilde{r} \sin\frac{\theta}{2}\right)\right]^2 \qquad (18)$$

It follows from this formula taking account of (10)

$$R = R_0 \frac{\pi(\hbar v_F)^2 N_i}{E_F^4}(U_0)^2 \int_0^\pi d\theta \sin^2\theta \left[\int_{\tilde{r}_1}^{\tilde{r}_2} d\tilde{r}\tilde{r}\, J_0\left(2\tilde{r} \sin\frac{\theta}{2}\right)\right]^2 \qquad (19)$$

This theoretical formula for resistance in case of the model potentials supported by annular well we compared with experimental curves from [5]. These two plots proved to be well consistent. Thus we can extract information about graphene sample parameters from this formula.

For this purpose let us consider the formula (19) in the vicinity of the Dirac point, i.e. for small Fermi energies. We have for small $E_F$: $2\tilde{r}\sin\frac{\theta}{2} = 2rk_F \sin\frac{\theta}{2} = 2r\frac{E_F}{\hbar v_F}\sin\frac{\theta}{2}$ is small.

Let us calculate the leading term in the asymptotic of (19) in the vicinity of the Dirac point. As soon as

$$J_0(x) \approx 1 - x^2/2 \ .$$

and

$$\int_0^\pi d\theta \sin^2\theta = \pi/2$$

we obtain

$$\frac{R}{R_0} = \frac{(\hbar v_F)^2 N_i}{4E_F^4}(U_0)^2 \pi^2 \left[\tilde{r}_2^2 - \tilde{r}_1^2\right]^2 \qquad (20)$$

Since (see above) $\tilde{r}_{1,2} = k_F r_{1,2} = \frac{E_F}{\hbar v_F} r_{1,2}$, we have

$$\frac{R}{R_0} = \frac{N_i}{4(\hbar v_F)^2} \left(S_{ring} U_0\right)^2 \tag{21}$$

As far as in the Dirac point the resistance has its maximal value, we can rewrite this formula as follows

$$\frac{R_{max}}{R_0} = N_i \left(S_{ring} \tilde{U}_0\right)^2, \quad \tilde{U}_0 = \frac{U_0}{2\hbar v_F}. \tag{22}$$

Thus, measuring the peak magnitude of the resistance one can find the unknown parameter $U_0 S_{ring}$ of the specimen.

See for comparison formula (7) for the delta-shell case.

### 3.Discussion

Above we have calculated the resistance in case of the potential model supported by annual well (see (19) and in the vicinity of the Dirac point see (22)). Let us compare these results with the formula (7) for resistance in case of the delta shell.

We shall consider the limit of (22) when $r_1 \to r_0$ and $r_2 \to r_0$. As soon as

$$S_{ring} \tilde{U}_0 = (r_1 + r_2)\pi(r_2 - r_1)\tilde{U}_0 = \pi(r_1 + r_2)(r_2 - r_1)\tilde{U}_0 \tag{23}$$

we have

$$\lim_{\substack{r_1 \to r_0 \\ r_2 \to r_0}} \left(S_{ring}\tilde{U}_0\right)^2 = \left(L_{circ}\bar{U}_0\right)^2 \tag{24}$$

Here we introduced

$$\bar{U}_0 = \tilde{U}_0 (r_2 - r_1) = \text{const}. \tag{25}$$

Substituting (24), (25) into (22) for the case, when the annual well is squeezing into a circumference with radius $r_0$ we obtain

$$\frac{R_{max}}{R_0} = N_i \left( L_{circ} \bar{U}_0 \right)^2 \tag{26}$$

This formula coincides with (7), where the role of $\tilde{V}_0$ plays $\bar{U}_0$. Thus we see that our new formula for the case of the model potential supported by the annual well tends to the old one (7) derived earlier in [1] for the case of the model potential equal to the delta function on the circumference. So our new formula tends to the delta shell case. Thus, in the limit of squeezing the annual well into a circumference we get delta shell potential of the form (see (2, 4)

$$V(r) = 2\hbar v_F \bar{U}_0 \delta(r - r_0). \tag{27}$$

Let us introduce the notion of intensity of a defect in annual well model

$$I_{ring}(U_0) = \left( S_{ring} \tilde{U}_0 \right)^2 \tag{28}$$

and intensity in $\delta$-shell model

$$I_{circ}(V_0) = \left( L_{circ} \tilde{V}_0 \right)^2 \tag{29}$$

Then

$$I_{circ}(U_0) = \left( L_{circ} \bar{U}_0 \right)^2 \tag{30}$$

So we have

$$I_{ring}(U_0) = I_{circ}(U_0) \tag{31}$$

We also introduce the notion of total intensity

$$I_{total} = N_i I_{ring}(U_0) = N_i I_{circ}(U_0) \tag{32}$$

Thus we have

$$\frac{R_{max}}{R_0} = I_{total}. \tag{33}$$

As soon as conductivity cannot be less than minimal value $\sigma_0$, we have $R_{max} \leq R_0$ and consequently

$$I_{total} \leq 1.$$

Using the derived formula (33) we can find $I_{total}$ from experimental data. Notice that this estimate works in the parameter range, where the Boltzmann equation is valid, i. e. not too near to the Dirac point.

## 4. Conclusions

We considered the high mobility graphene resistance basing on two models of the short-range scattering potential: the delta-shell potential and the annual well one. An advantage of the former is simplicity of analytics, while an advantage of the latter is its regularity. Both of the models give qualitatively similar behavior as a function of the Fermi energy. Quantitative correspondence of both parameters is established. Characteristic parameters of the scattering efficiency for a single defect and for a specimen with randomly distributed defects are introduced. These parameters can be found from experiments.

## References

1. S.A. Ktitorov, N.E. Firsova, Phys. Solid State, **61**, 609 (2019).
2. N. E. Firsova, Nanosystems: Physics, Chemistry, Mathematics **4,** 538 (2013).
3. N.E. Firsova, S.A. Ktitorov, Phys. Lett. A, **374**, 1270 (2010).
4. D.S.Novikov. Phys.Rev.B76,245(2007)
5. K.I. Bolotin, K.I. Sikesh, Z. Jiang et al., Solid State Communications (2008) **146**, 351.
6. S.A. Ktitorov, N.E. Firsova, Phys. Solid State **53**: 411 (2011).